\documentclass[a4paper,twoside,10pt]{article}
\usepackage{graphicx}

\begin{document}
\begin{titlepage}
\begin{center}
\bf{Impact crater formation:a simple application of solid state physics}
\end{center}

\begin{center}
V. Celebonovic$^1$ and J.Souchay$^2$
\end{center}

\begin{center}
$^1$Institute of Physics, Pregrevica 118, 11080 Zemun-Belgrade,Serbia
\end{center}

\begin{center}
email: vladan@ipb.ac.rs
\end{center}

\begin{center}
$^2$SYRTE,Observatoire de Paris,61 Av.de l'Observatoire,75014 Paris,France
\end{center}

\begin{center}
email:Jean.Souchay@obspm.fr
\end{center}

\begin{abstract}
This contribution is a first step aiming to address a general question: what can be concluded on impact craters which exist on various planetary system objects, by combining astronomical data and known theoretical results from solid state physics. Assuming that the material of the target body is of crystaline structure,it is shown that a simple calculation gives the possibility of estimating the speed of the impactor responsible for the creation of a crater.A test value,calculated using observed data on the composition of some asteroids,gives a value of the speed in good agreement with results of celestial mechanics.
\end{abstract}
\footnote{Presented at the 6 SREAC meeting in Belgrade (Serbia),September 2009.,and accepted for the proceedings}
\end{titlepage}
\section{Introduction}

Craters of various sizes exist on planets,their satellites and (at least some of) the asteroids. Their existence is a consequence of two kinds of processes: volcanic eruptions in the interior or impacts  of external bodies into the surface of the target body. The existence of impact craters on the surfaces of various objects testifies about the density of small bodies in the early solar system. For an example of a recent study see [4]. An impact of a sufficiently big body on the surface of the Earth could potentially have catastrophic consequences for life on Earth. That is the "practical" motivation for studies of impact craters. 

Bombardement of surfaces of the inner planets occured $3.8$-$3.9$ billion years ago [4]. As only the consequences of this event,in the form of craters are observable at present, the general question is what can be concluded on the impactors which formed the craters, by using astronomical observation and results of experimental and/or theoretical solid state physics.

The aim of this paper is to present a simple calculation which gives the possibility of estimating the speed of the impactor in terms of various material parameters of the target. 

The calculation to be discussed is based on a simple physical idea: the formation of a crater will be possible if the kinetic energy per unit volume of an impactor in the moment of impact into the surface of a target is equal to or greater than the lattice energy per unit volume of the target.
  
\section{Calculation}

It can be shown in solid state physics [2] that the energy per unit volume of  crystal lattice is given by

\begin{equation}
	\frac{E}{V}= \frac{\pi^{2}}{10} \frac{(k_{B}T)^{4}}{(\hbar \overline{V})^{3}}
\end{equation}

The symbols $k_{B}$ and $\hbar$ denote,respectively, the Boltzmann constant and Planck's constant; $T$ is the temperature and $\overline {V}$ is the speed of elastic waves in the material of the target. 

It can be shown in statistical physics that $\overline {V}$  is related to the pressure $P$ and density $\rho$ of the material by the following simple relation:
\begin{equation}
\overline{V} = \frac{\partial P}{\partial \rho}	
\end{equation}
 
The kinetic energy per unit volume of the impactor is:
\begin{equation}
	\frac{E}{V} = \frac{1}{2} \rho_{1} v^{2}
\end{equation}

where $\rho_{1}$ is the mass density of the impactor and $v$ is its speed. Equating the right-hand sides of Eqs.(1) and (3),simple algebra gives the following result for the speed of the impactor:

\begin{equation}
v^{2}=\frac{\pi^2}{5\rho_{1}} \frac{(k_{B}T)^{4}}{(\hbar \overline{V})^{3}}
\end{equation}

Inserting Eq.(2) into Eq.(4) the speed of the impactor can be expressed in a more general form

\begin{equation}
v^{2}=\frac{\pi^2}{5\rho_{1}} \frac{(k_{B} T)^{4}}{\hbar^{3}} \left(\frac{\partial P}{\partial \rho}\right)^{-3/2} 
\end{equation}
This relation gives the speed which  the impactor made up of a material having mass density $\rho_{1}$ hitting a target made up of material having pressure $P$ and density $\rho$ on temperature $T$ must have in order to make an impact crater in it. In order to render it applicable,and thus comparable with estimates obtained by celestial mechanics, one must either insert "`by hand" the mean value of the derivative ${\partial P}/{\partial \rho}$ or choose an appropriate equation of state for the material of the target.

\section{A test example}

In order to get to know the mean mass density of an incoming asteroid (or comet),one has at first to determine its chemical composition. This can be done by ground based observation , such as [1], but also from space probes (a recent example is [5]). Once the surface chemical composition is known,assuming that the bulk composition of the object is similar to the observable surface composition, the mean mass density can be determined from laboratory data.
It was recently shown [3] by analysis of $IR$ spectra of four $S$ type asteroids that they contain variable amounts of minerals similar to olivine on their surfaces. The chemical formula of olivine is $(Mg,Fe)_{2}SiO_{4}$ and its density is approximately $3300$ $kg/m^{3}$. The temperature of the target can be taken as $T=300$ $K$, and the mean value of the speed of elastic waves as $3 km/s$. This is the mean observed value of the speed of seismic waves. 

Inserting all these data into Eq.(4) gives $v\cong 16.3$ $km/s$. 
Physically speaking,this is the value of the speed that an object made up entirely of olivine should have in order to make an impact crater on the surface of the Earth.

This value of the impactor velocity was calculated by using basic principles of solid state physics,while it is normally considered as a problem in celestial mechanics. It was thus recently shown that the impact velocity of asteroid $99942$ Apophis is $12.6$ $km/s$ (http://neo.jpl.nasa.gov/risk/a99942.html). A somewhat larger value of $20.3$ $km/s$ was quoted in (http://neo.jpl.nasa.gov/news/news

165.html) as the average impact velocity of $NEAs$. Although one of these values is lower,and the other one higher than the value of the impact speed calculated in this contribution, the present result is clearly of the same order. It is enocuraging for the basic idea of the present paper that two completely different methods,celestial mechanics and solid state physics, give values which are mutually close.

\section{Discussion}

The calculation presented in this contribution is mathematically simple,and it is founded on just one physical idea. Namely,it is assumed that if an impacting body is to make a crater in a target,its kinetic energy per unit volume has to be equal to the lattice energy per unit volume of the material of the target. This assumption has its inherent physical limitations.

Namely,in any collision,the kinetic energy of the incoming projectile is shared between the projectile and the target,and it is partially used for heating of the target in the immediate vicinity of the point of impact. The quantity of heat which is released in a collision depends obviously on the kinetic energy of the impactor and the heat capacity of the target. Not taking the localised heating of the target into account implies a limitation to the case of low velocity impacts in which the conversion of a part of the kinetic energy of the impactor in heat energy can be neglected. 

Determining the heat capacity of any material is an interesting task. In a solid with lattice structure it depends on the phononic frequency distribution function,which in turn depends on the composition of the material.

A continuation of the calculation reported in this contribution is planned, but with the introduction of a more elaborate energy balance of impacts. This practically means that heating of a target in an impact will be taken into account. Another interesting aspect of impacts is the propagation of the thermal energy released in an impact,and the possible consequences it may have on the target.

Qualitatively speaking,the impact is an instantenious ( or very short lived ) event. This implies that a part of the kinetic energy of the impactor is converted into heat in a very short time interval,and a further implication is that a heat wave with decaying amplitude propagates through the target from the point of impact. Now immagine a situation of a high speed impact in a target made up of a weakly bound material with low heat capacity. Such an impact could easily lead to a break-up of the target. 

Rigorous work along the lines of this qualitative discussion is at present being planned.          

\section{Note added}

A further illustration of the applicability of the ideas developed in this contribution to asteroid impacts is provided by the asteroid $2007VK184$,the impact speed of which has been estimated as 19.19 $km/s$. Detailed results on the possible impacts of this objects are avaliable on the following address:  http://neo.jpl.nasa.gov/risk/2007vk184.html . 

\section{Acknowledgement}

This contribution was prepared within the project $141007$ financed by the Ministry of Science and Technological Development of Serbia.







{}

\end{document}